\documentclass[aps,prl,preprint,tightenlines,superscriptaddress,showpacs,byrevtex]{revtex4}
\usepackage{graphicx} 
\usepackage{dcolumn}  

\graphicspath{{ps}}

\begin{document}


\preprint{\vbox{ \hbox{   }
                 \hbox{BELLE-CONF-0474}
                 \hbox{ICHEP04 8-0673} 
}}

\title{ \quad\\[0.5cm]  Moments of the Electron Energy Spectrum in Semileptonic $B$ Decays at Belle}

\affiliation{Aomori University, Aomori}
\affiliation{Budker Institute of Nuclear Physics, Novosibirsk}
\affiliation{Chiba University, Chiba}
\affiliation{Chonnam National University, Kwangju}
\affiliation{Chuo University, Tokyo}
\affiliation{University of Cincinnati, Cincinnati, Ohio 45221}
\affiliation{University of Frankfurt, Frankfurt}
\affiliation{Gyeongsang National University, Chinju}
\affiliation{University of Hawaii, Honolulu, Hawaii 96822}
\affiliation{High Energy Accelerator Research Organization (KEK), Tsukuba}
\affiliation{Hiroshima Institute of Technology, Hiroshima}
\affiliation{Institute of High Energy Physics, Chinese Academy of Sciences, Beijing}
\affiliation{Institute of High Energy Physics, Vienna}
\affiliation{Institute for Theoretical and Experimental Physics, Moscow}
\affiliation{J. Stefan Institute, Ljubljana}
\affiliation{Kanagawa University, Yokohama}
\affiliation{Korea University, Seoul}
\affiliation{Kyoto University, Kyoto}
\affiliation{Kyungpook National University, Taegu}
\affiliation{Swiss Federal Institute of Technology of Lausanne, EPFL, Lausanne}
\affiliation{University of Ljubljana, Ljubljana}
\affiliation{University of Maribor, Maribor}
\affiliation{University of Melbourne, Victoria}
\affiliation{Nagoya University, Nagoya}
\affiliation{Nara Women's University, Nara}
\affiliation{National Central University, Chung-li}
\affiliation{National Kaohsiung Normal University, Kaohsiung}
\affiliation{National United University, Miao Li}
\affiliation{Department of Physics, National Taiwan University, Taipei}
\affiliation{H. Niewodniczanski Institute of Nuclear Physics, Krakow}
\affiliation{Nihon Dental College, Niigata}
\affiliation{Niigata University, Niigata}
\affiliation{Osaka City University, Osaka}
\affiliation{Osaka University, Osaka}
\affiliation{Panjab University, Chandigarh}
\affiliation{Peking University, Beijing}
\affiliation{Princeton University, Princeton, New Jersey 08545}
\affiliation{RIKEN BNL Research Center, Upton, New York 11973}
\affiliation{Saga University, Saga}
\affiliation{University of Science and Technology of China, Hefei}
\affiliation{Seoul National University, Seoul}
\affiliation{Sungkyunkwan University, Suwon}
\affiliation{University of Sydney, Sydney NSW}
\affiliation{Tata Institute of Fundamental Research, Bombay}
\affiliation{Toho University, Funabashi}
\affiliation{Tohoku Gakuin University, Tagajo}
\affiliation{Tohoku University, Sendai}
\affiliation{Department of Physics, University of Tokyo, Tokyo}
\affiliation{Tokyo Institute of Technology, Tokyo}
\affiliation{Tokyo Metropolitan University, Tokyo}
\affiliation{Tokyo University of Agriculture and Technology, Tokyo}
\affiliation{Toyama National College of Maritime Technology, Toyama}
\affiliation{University of Tsukuba, Tsukuba}
\affiliation{Utkal University, Bhubaneswer}
\affiliation{Virginia Polytechnic Institute and State University, Blacksburg, Virginia 24061}
\affiliation{Yonsei University, Seoul}
  \author{K.~Abe}\affiliation{High Energy Accelerator Research Organization (KEK), Tsukuba} 
  \author{K.~Abe}\affiliation{Tohoku Gakuin University, Tagajo} 
  \author{N.~Abe}\affiliation{Tokyo Institute of Technology, Tokyo} 
  \author{I.~Adachi}\affiliation{High Energy Accelerator Research Organization (KEK), Tsukuba} 
  \author{H.~Aihara}\affiliation{Department of Physics, University of Tokyo, Tokyo} 
  \author{M.~Akatsu}\affiliation{Nagoya University, Nagoya} 
  \author{Y.~Asano}\affiliation{University of Tsukuba, Tsukuba} 
  \author{T.~Aso}\affiliation{Toyama National College of Maritime Technology, Toyama} 
  \author{V.~Aulchenko}\affiliation{Budker Institute of Nuclear Physics, Novosibirsk} 
  \author{T.~Aushev}\affiliation{Institute for Theoretical and Experimental Physics, Moscow} 
% ...
\collaboration{The Belle Collaboration}

\noaffiliation

\begin{abstract}
We report a measurement of the inclusive electron energy spectrum for semileptonic decays of $B$ mesons in a  $140\,{\rm fb}^{-1}$ data sample collected near the $\Upsilon(4S)$ resonance,
with the Belle detector at the KEKB asymmetric energy $e^+ e^-$ collider.  We determine the first and second moments of the spectrum for threshold values of the electron energy between 0.6 and 1.5 GeV.
\end{abstract}


\maketitle

\tighten

{\renewcommand{\thefootnote}{\fnsymbol{footnote}}}
\setcounter{footnote}{0}

\section{Introduction}

The Cabibbo-Kobayashi-Maskawa matrix element $|V_{cb}|$ can be extracted from the inclusive
branching fraction for semileptonic $b$ hadron decays ${ \hbox{B}} ( B \to X_c \ell \nu )$\cite{bigi,wise}.  Several studies have shown that the spectator model decay rate is the leading term in a well-defined expansion controlled by 
the parameter $\Lambda _{\rm QCD}/m_b$. Non-perturbative corrections
to this leading approximation arise only at order $1/m_b^2$ and above. The key issue in this approach is the ability to separate non-perturbative corrections, which can be expressed as a series in powers of $1/m_b$, and perturbative corrections, expressed in powers of 
$\alpha _s$.

The coefficients of the $1/m_b$ power terms are expectation values of operators that include non-perturbative physics.
Two different expansions exist, reflecting a difference in the
approach used to handle the energy scale $\mu$ that separates long-distance from short-distance physics.
In this note, we use the short distant mass expansion
that defines  the non-perturbative operators using a mass scale $\mu \approx 1$ GeV \cite{ref:1}.

The shape of the lepton spectrum provides constraints on the heavy
quark expansion based on local Operator Product Expansion (OPE), which
calculates properties of the $ B \to X_c \ell \nu $
transitions.  So far, measurements of the electron energy distribution
have been made by the DELPHI, CLEO, and BABAR collaborations
\cite{ref:4,cleoel,babarel}.  In this note we report a
measurement of the first and second moment of the electron energy
spectrum with a minimum electron momentum cut ranging between 0.6 and
1.5 GeV in the $B$ meson rest frame.  We measure, independently, the 
electron energy moments for the semileptonic decays of the
$B^+$ and the $B^0$ mesons~\cite{CC}.  If quark hadron duality applies in $ B
\to X_c \ell \nu $ decays then the moments from the $B^+$
and $B^0$ should agree.
Up to now such a measurement has been impossible to perform. However,
in this measurement we find good agreement between the moments.

The data used in this analysis was collected with the Belle detector
at the KEKB~\cite{KEKB} asymmetric energy $e^+ e^-$ collider.
The Belle~\cite{Belle} detector is a large-solid-angle magnetic spectrometer that
consists of a three-layer silicon vertex detector (SVD), a 50-layer central drift chamber (CDC), 
an array of aerogel threshold \v{C}erenkov counters (ACC), 
a barrel-like arrangement of time-of-flight scintillation counters (TOF), and an electromagnetic calorimeter comprised of CsI(Tl) crystals (ECL) located inside a super-conducting solenoid coil that provides a 1.5~T magnetic field.  An iron flux-return located outside of
the coil is instrumented to detect $K_L^0$ mesons and to identify muons (KLM).

Events are selected by  fully reconstructing one of the $B$ mesons, produced in pairs from $\Upsilon (4S)$, in several hadronic decay modes.  Prompt semileptonic decays ($b \rightarrow x \ell \nu$) of the non-tag side $B$ mesons are separated from cascade charm decays ($b \rightarrow c \rightarrow y \ell \nu$), based on the correlation between the flavour of the tagged $B$ and the lepton charge.

The present results are based on a $140\,{\rm fb}^{-1}$ data sample collected at the $\Upsilon (4S)$ resonance, which contains $1.5 \times 10^8$ $B \overline B$ pairs.  An additional $15\,{\rm fb}^{-1}$ data sample taken at 60~MeV below the $\Upsilon (4S)$ resonance (off$-$resonance) is used to perform background subtraction from the 
$e^+e^- \rightarrow q \overline q$ process.

We use a fully simulated generic Monte Carlo sample generated with
the qq98 event generator \cite{BelleMC}.  This sample is equivalent to
about three times the beam luminosity of the real data. Simulated
events are required to satisfy tight hadron selection criteria.

\section{Full Reconstruction Sample}


The hadronic decay of the tag-side $B$ meson is fully reconstructed in
the decay modes $B \to D^{(*)} \pi^+, D^{(*)} \rho^+, D^{(*)} a_1^+$,
yielding a high purity $B$ meson sample. The following sub-decay modes are
implemented:
\begin{itemize}
\item $D^{*0}\to D^0\pi^0, D^0\gamma$,
\item $D^{*+}\to D^0\pi^+, D^+\pi^0$,
\item $D^0\to K\pi, K\pi\pi^0, K\pi\pi\pi, K_S\pi\pi, K_S\pi^0$ and
\item $D^+\to K\pi\pi, K_S\pi$.
\end{itemize}
If multiple candidates are found in one event, we select the best candidate
based on $\Delta E$ and other variables. There are two conditions that affect the best
candidate selection.  The candidate may be
reconstructed from at least one particle from the other side $B$, or
the final state particles may be incorrectly reconstructed.  

\section{Electron Selection}

We identify electrons produced  in semileptonic $B$ decays on the
``non-tag'' side.   
Low-momentum particles spiral in the CDC~detector. If the particles
pass close to the interaction point  more than once, they can lead to
multiple reconstructed tracks. Those tracks are removed. 
The remaining tracks are required to satisfy tight restrictions on the
impact parameters.
In addition, we require the
tracks to be in the barrel region of the detector, corresponding to an
angular acceptance of $35^{\circ} \leq \theta_{\mathrm{lab}} \leq
125^{\circ}$, where $\theta_{\mathrm{lab}}$ is the polar angle of the
track relative to the $z$ axis (opposite the positron beam line).


Tracks that pass the above selection criteria and are not used in the
reconstruction of the tag-side $B$ meson are considered as electron
candidates. Electron identification is based on a
combination of specific ionisation $dE/dx$ measurements
in the CDC, the response of the ACC, the shower shape in the ECL, and the
ratio of energy deposit in the ECL to the momentum measured by the
tracking system ($E/p$) \cite{Belle}.   The electron identification
efficiencies include the detector
acceptance as well as the selection efficiency. The errors for the tracking and
electron detection efficiencies are $\pm 1\%$ and $\pm 2\%$,
respectively. Electron identification efficiency is determined by analysing
a data sample where simulated single electron tracks are overlaid on
hadronic events taken from real experimental data. The electron
momentum spectrum is corrected by using this momentum dependent
electron detection efficiency.

The momentum of electron candidates is measured in the $B$ meson rest
frame ($p^{*(B)}$), exploiting the knowledge on the momentum of the
fully reconstructed $B$. To ensure a good electron identification
performance we require $p^{*(B)}_{e} \geq 0.6$ GeV/$c$.

Backgrounds from $J / \psi$, $\psi (2S)$, photon conversions in the
material of the detector, and Dalitz decays from $\pi^0$ and $\eta$
mesons are reduced by imposing veto cuts.  We look for additional electrons
in the event, which satisfy looser identification criteria and $p_{e}
\geq 0.3$ GeV/$c$.  We then calculate invariant masses ($m_{ee}$) for each
electron candidate when combined with an opposite-charge electron that
passes the loose selection. In the Dalitz case  we require an
additional photon ($m_{ee\gamma}$). We reject the electron if
$m_{ee}$ lies within the nominal $J / \psi$ and $\psi (2S)$ mass
regions,  or $m_{ee\gamma}$ lies within the nominal $\pi^0$ and $\eta$
mass regions.  To remove photon conversions we reject candidates by placing cuts on the vertex of the decay and $m_{ee}$ less than 100 MeV/$c^2$.




\section{Number of Tagged Events}
 
We select events containing a fully reconstructed $B$ candidate on one side and 
an electron candidate on the other side.

For each selected event, we calculate the beam-constrained mass $M_{\mathrm{bc}}$ and
$\Delta E$:
\begin{equation}
  M_{\mathrm{bc}} = \sqrt{(E^*_{\mathrm{beam}})^2-(\vec p^*_B)^2}, \quad \Delta E =
  E^*_B-E^*_{\mathrm{beam}},
\end{equation}
where $E^*_{\mathrm{beam}}$, $\vec p^*_B$ and $E^*_B$ are the beam energy, the
$B$ 3-momentum and the $B$ energy in the centre of mass
frame, respectively. 
Events
with   $M_{\mathrm{bc}}>5.27~\mathrm{GeV/}c^2$  and $-0.06~\mathrm{GeV}<\Delta E< 0.08~\mathrm{GeV}$ are considered to be signal candidates.

\begin{figure}[htb]
  \begin{tabular}{cc}
\includegraphics[width=0.48\textwidth]{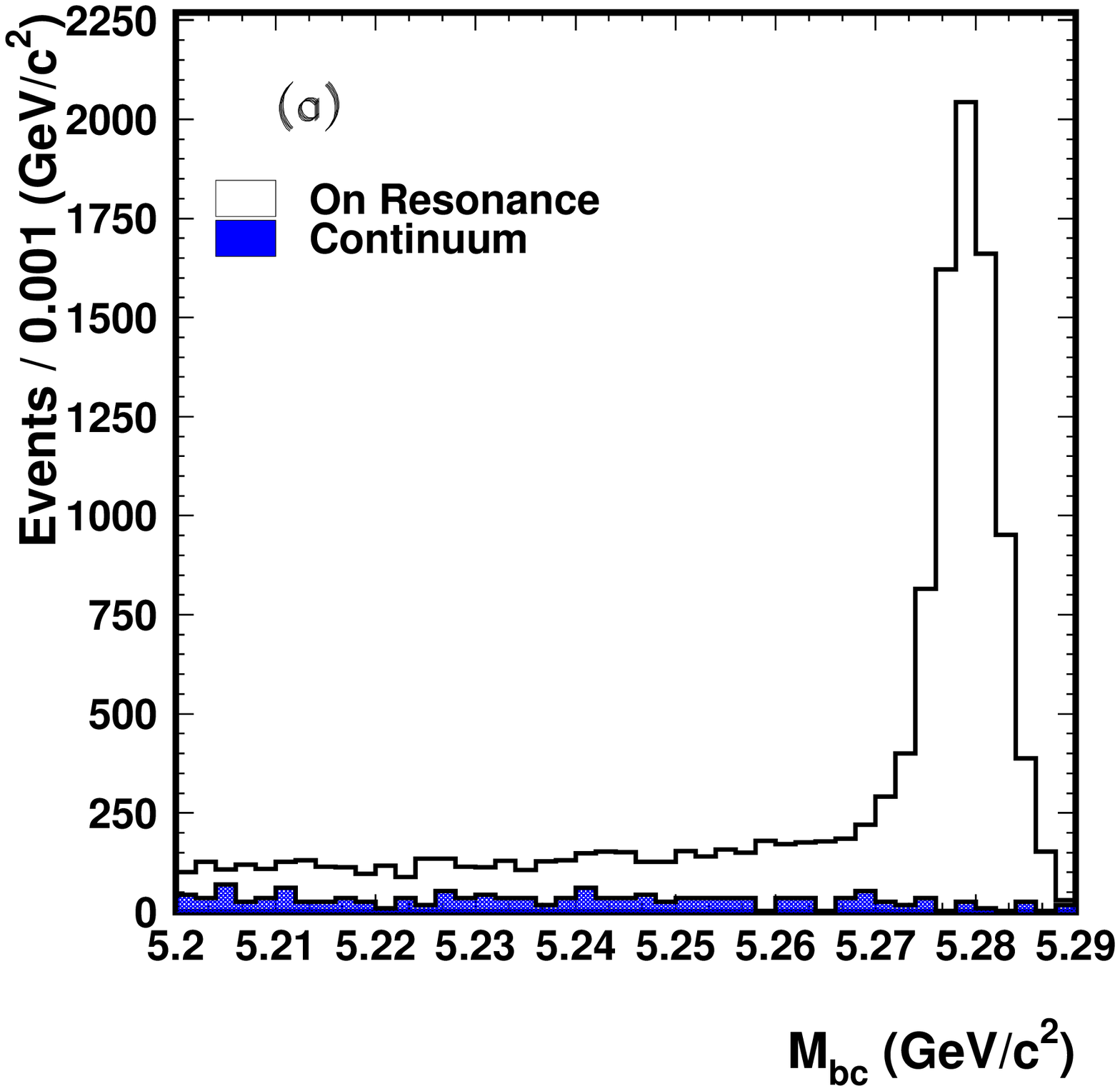} &\includegraphics[width=0.48\textwidth]{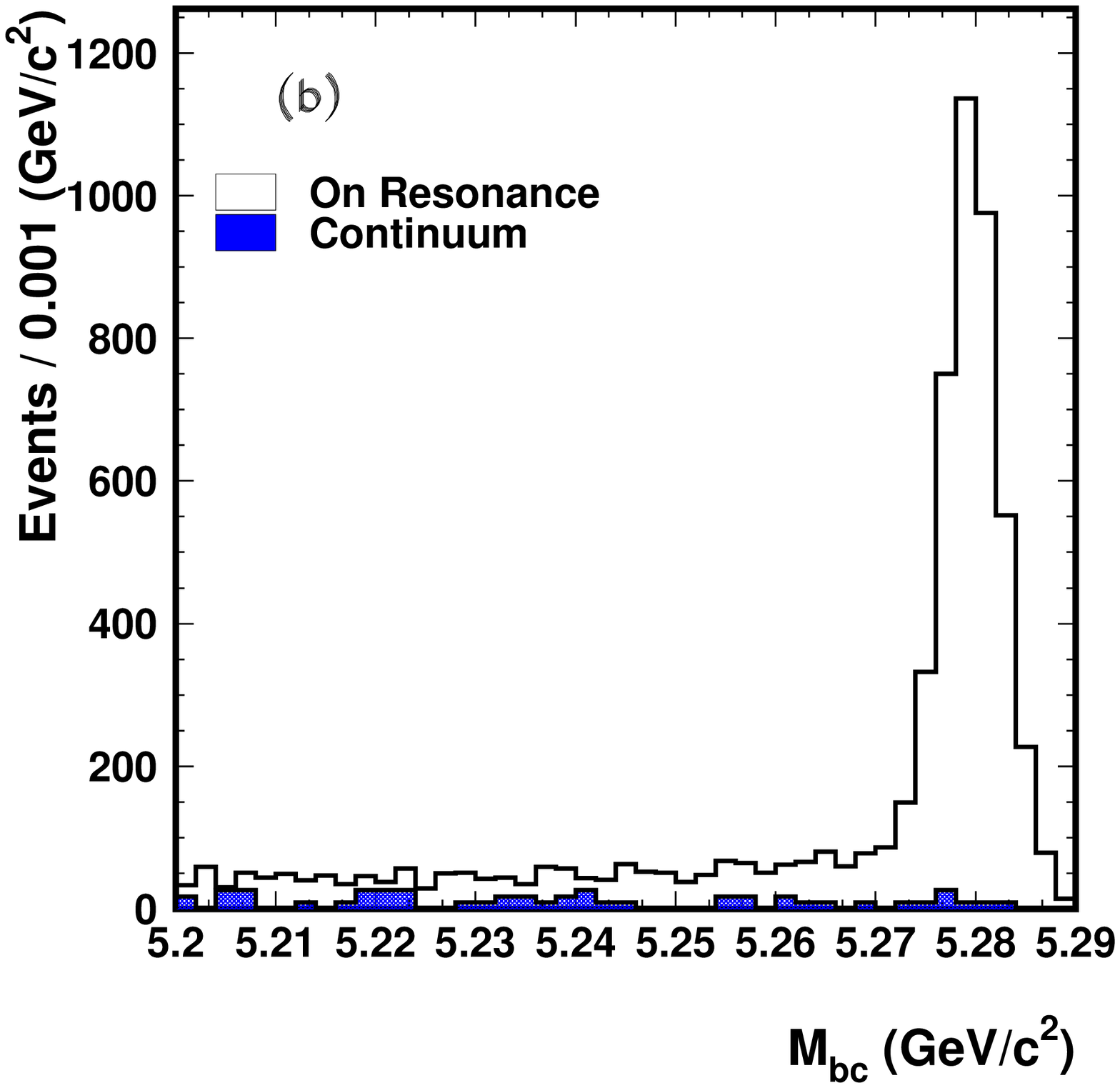}\\
\includegraphics[width=0.48\textwidth]{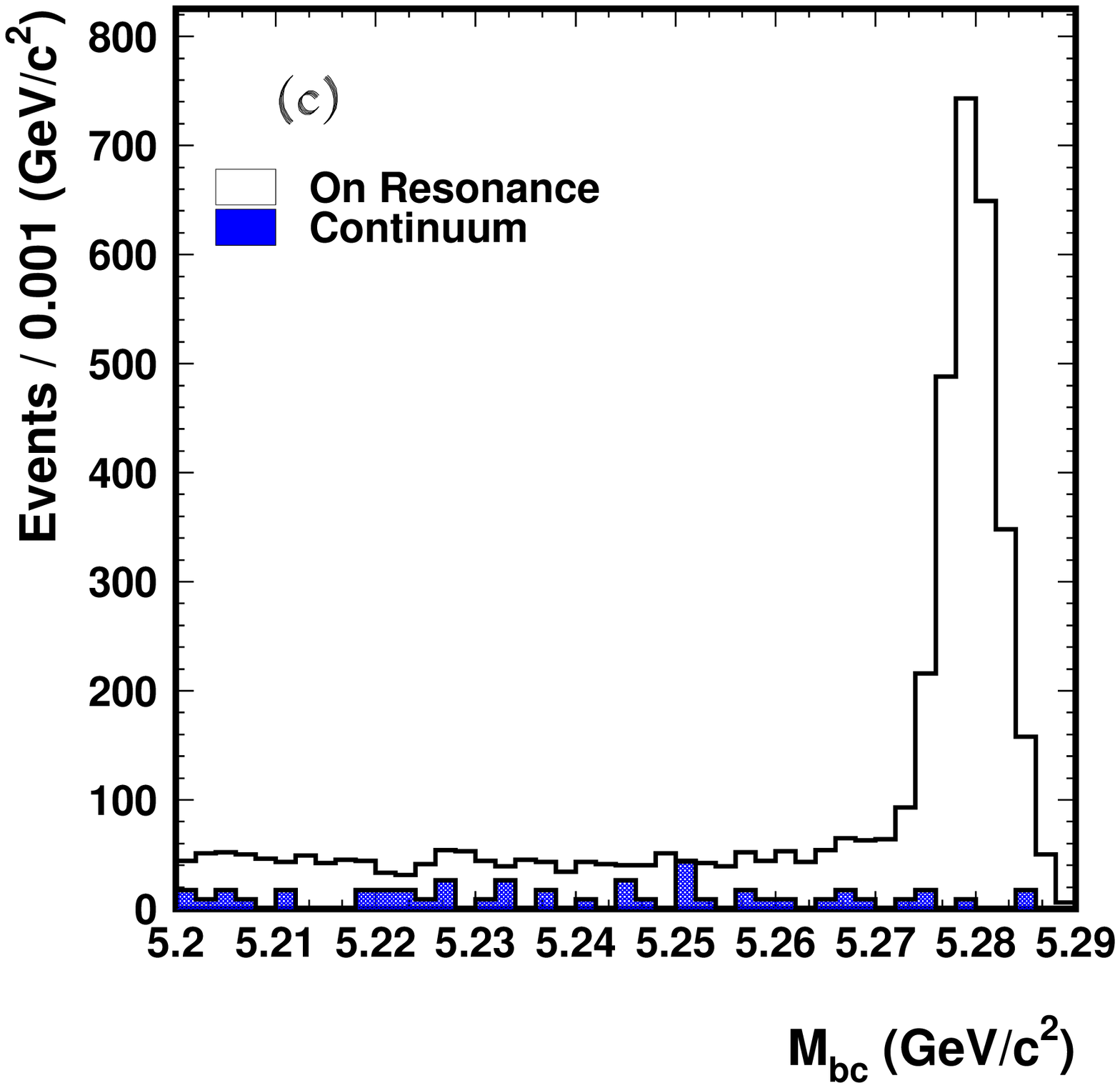}&\\
\end{tabular}
\caption{The beam constrained mass after electron selection cuts and
  $\Delta E$ cuts for the $B^+$ right sign electron sample~(a),
  $B^0$ right sign electron sample~(b) and the $B^0$ wrong sign electron sample~(c). }
\label{mbc}
\end{figure}

Each electron is assigned to the ``right sign'' sample if the electron
has the opposite flavour to the fully reconstructed $B$, and to the
``wrong sign'' sample if the electron has the same flavour as the fully
reconstructed $B$.  In events without $B^0 \overline{B}{}^0$ mixing, the
primary electrons belong to the ``right sign'' sample while the
secondary electrons contribute to the ``wrong sign'' sample.
Figure \ref{mbc} shows the beam constrained mass distribution for $B^0$ and $B^+$ right sign and $B^0$ wrong sign electrons, with scaled off resonance background overlaid. 
The number of events in the  signal region, is $3367 \pm 58$ ($1659
\pm 40 $) and 
$6831 \pm 83$ for the $B^0$ right sign (wrong sign) and $B^+$ right sign candidates, respectively.

\section{Background Subtraction}

The selected electron energy spectrum is contaminated by background processes, which should be evaluated and subtracted from the distribution before the extraction of the moments. The residual background is from:
\begin{itemize}
 \item continuum background,
\item full-reconstruction background,
\item background from secondary decays,
\item $J/ \psi, \psi (2S)$, Dalitz decays and photon conversions where
  one electron of the pair has escaped detection and
\item fake electrons.
\end{itemize}

The electron momentum spectrum of the continuum background is derived
from off$-$resonance
data and is normalised using the off$-$ to on$-$resonance luminosity
ratio.  To account for the low statistics in the off$-$resonance data we
fit an exponential to the electron energy distribution.  We fit only the
$B^+$ candidates then apply the same continuum cut to the $B^0$,
scaled to the electron yield of the wrong
sign and right sign $B^0$ samples.
There are insufficient statistics to fit the off$-$resonance
$B^0$ samples.

The electron momentum spectrum of the combinatorial background
is derived from the generic $B \overline B$ Monte Carlo events where
either reconstruction or flavour assignment of  the tagged $B$ meson
is not carried out correctly.  The $B \overline B$ Monte Carlo events are
normalised to the on resonance data $M_{\mathrm{bc}}$ side band ($M_{\mathrm{bc}}<5.25~\mathrm{GeV/}c^2$) after continuum background subtraction in this region.



We also correct for cases where the fully reconstructed
$B$ is correctly tagged, but the electron does not directly originate
from a $B$ decay or is a hadron misidentified as an electron. These
backgrounds are irreducible; to estimate the magnitude of the
contributions we normalise the $B \overline B$ Monte Carlo to the
electron yield after continuum and combinatorial background
subtraction.

Secondary electrons arising from $B \to D^{(*)} \to e$, $B \to
\tau \to e$, and $B \to D^{(*)} \to \tau \to e$ decays are simulated
using the latest published branching fractions~\cite{PDG}.


Contributions from $J/ \psi, \psi (2S)$ decays,
photon conversions, and Dalitz decays are insignificant after the electron
selection cuts.  The surviving backgrounds, where one electron of the
pair has escaped detection, are estimated by the Monte Carlo
simulation and subtracted along with the major secondary
backgrounds.

The Monte Carlo yield due to fakes is corrected for the difference in
the fake rate for data and Monte Carlo measured with samples of $
K^{0}_{S} \to \pi^+\pi^-$ decays. The magnitude of the correction is
assigned as the systematic uncertainty on the fake yield.

\begin{figure}[htb]
  \begin{tabular}{cc}
    \includegraphics[width=0.48\textwidth]{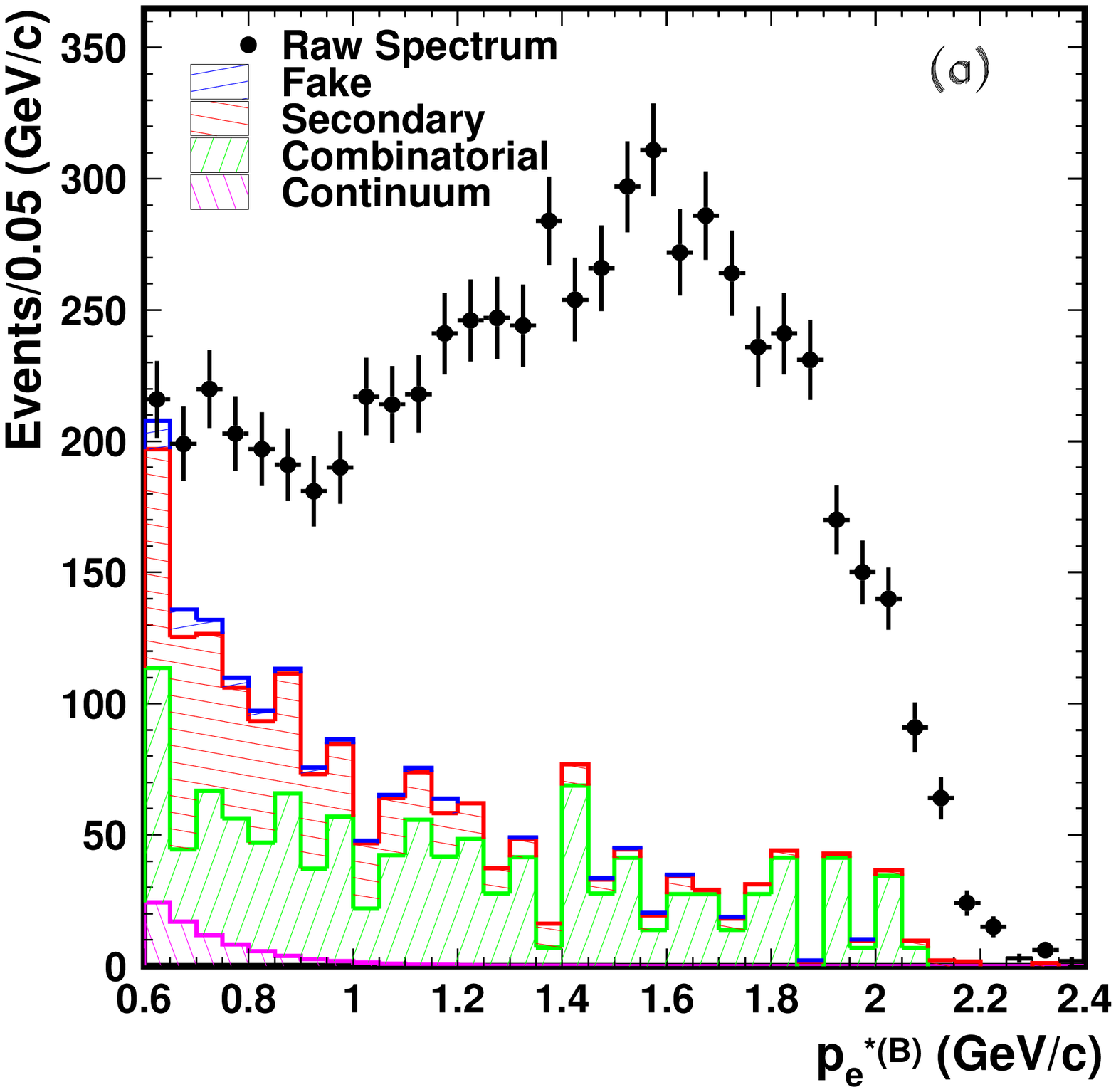}
 & \includegraphics[width=0.48\textwidth]{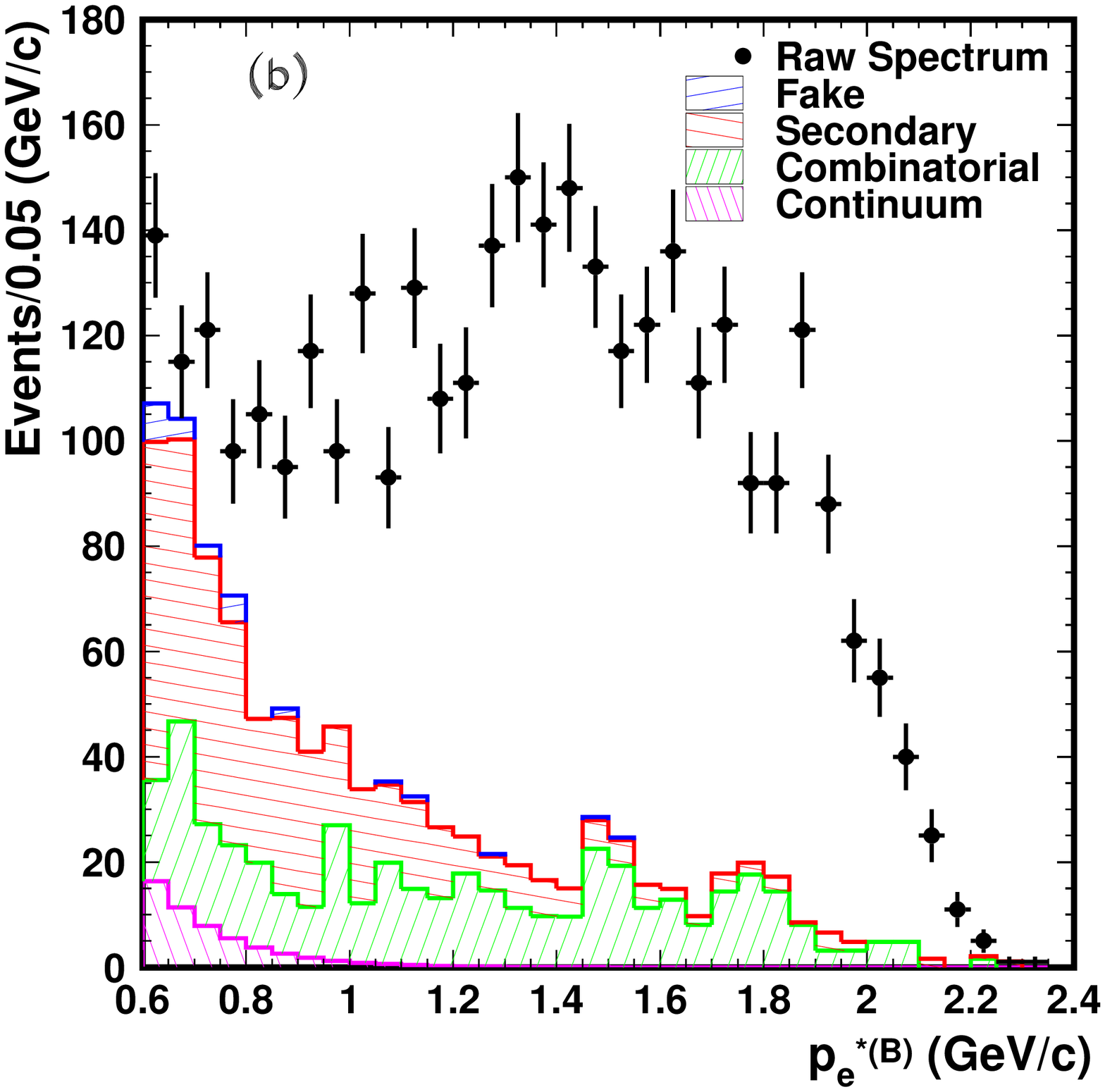}\\
\includegraphics[width=0.48\textwidth]{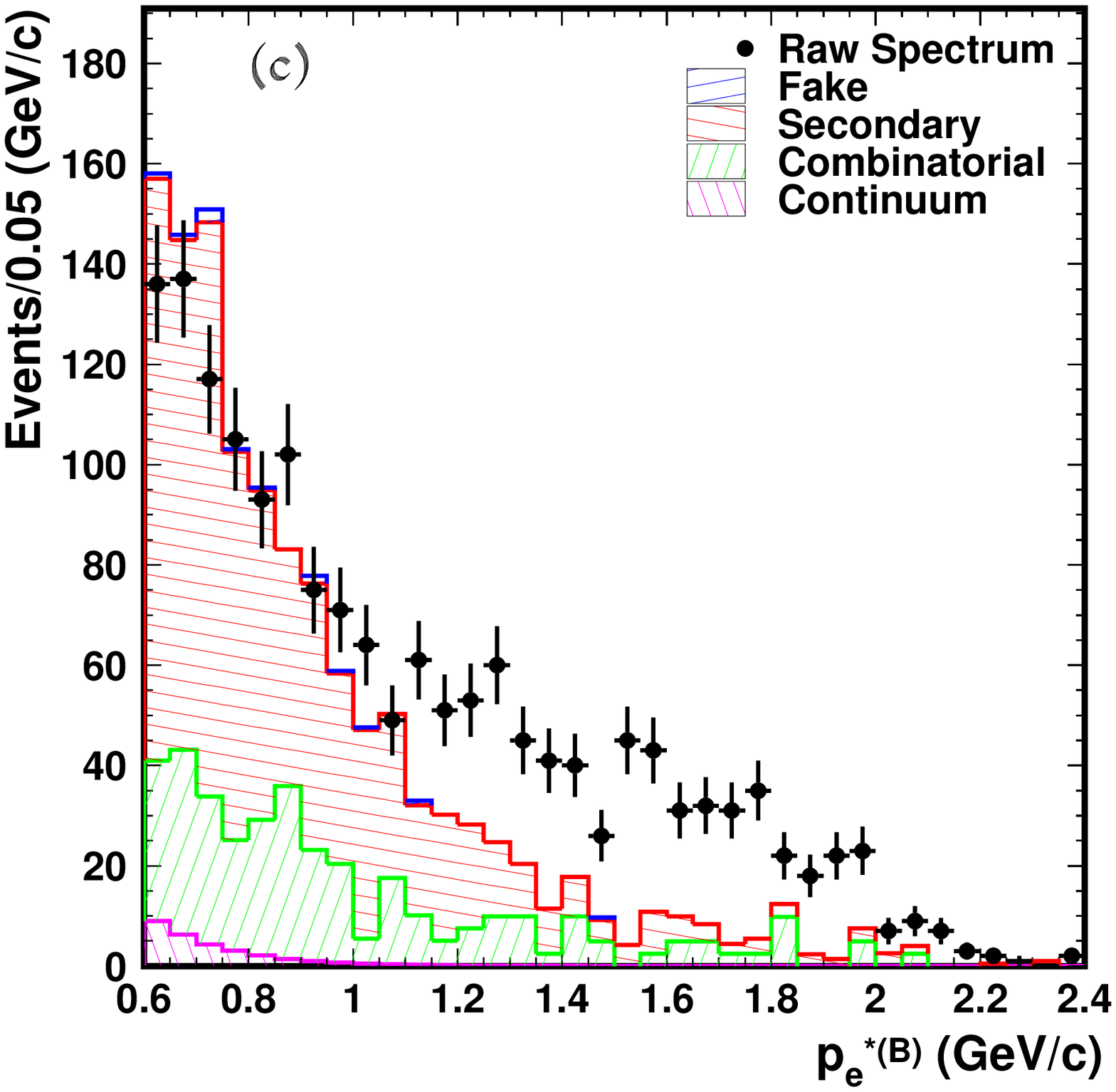} & \\
\end{tabular}
\caption{Breakdown of the backgrounds in the electron momentum spectra
  for $B^+$ right sign~(a), $B^0$ right sign~(b) and $B^0$ wrong sign~(c) electrons.  }
\label{rawwithbg}
\end{figure}

Figure \ref{rawwithbg} shows the raw electron momentum spectrum
with all background contributions overlaid.
Table \ref{bzero} summarises the number of detected electrons
and the contributions from these backgrounds.

\section{Mixing Corrections}
To reduce the contamination from  cascade semileptonic charm decays,
we determine whether the electron's parent and the fully reconstructed
$B$ have the same or opposite flavour.  For events tagged as $B^+$ we require
the right sign tag. For events tagged as $B^0$, we
introduce a dependence on $B^0 \overline{B}{}^0$ mixing which is  taken
into account by solving the equations; $N_{\mathrm{right}} = N_{\mathrm{p}} ( 1-\chi_d ) + N_{\mathrm{c}} \chi_d$,
$N_{\mathrm{wrong}}=N_{\mathrm{p}} \chi_d + N_{\mathrm{c}} ( 1 - \chi_d)$ where $N_{\mathrm{right}}$ and
$N_{\mathrm{wrong}}$ are the number of electrons with the right and wrong
signs, respectively. $N_{\mathrm{p}}$ and $N_{\mathrm{c}}$ are the number of electrons
from the prompt and cascade semileptonic decays, respectively and
$\chi_d = 0.186 \pm 0.004$ \cite{PDG} is the mixing probability.

\section{Corrections to the Electron Energy Spectrum}
The electron energy spectrum is generated via Monte Carlo simulation
of $B \rightarrow X_c e \nu$ decays using the qq98 event generator
\cite{BelleMC}.   The spectrum from $B \rightarrow X_c e \nu$ is
modelled using four components: $X_c = D$ (ISGW2~\cite{ref:8}), $D^*$
(HQET~\cite{ref:7}), higher resonance charm meson states
$D^{**}$(ISGW2) and non-resonant $D^{(*)} \pi$ (Goity and
Roberts~\cite{ref:9}).   To account for the most recent theoretical
and experimental results, we re-weight the $D$ and $D^*$
components in $p^{*(B)}$ to the spectra generated with current world
average form factors \cite{PDG}.

The available Monte Carlo sample does not incorporate ${\mathcal O}(\alpha)$ QED
corrections.  We correct the background
subtracted electron spectrum with energy-dependent weights using the
PHOTOS \cite{PHOTOS} package.

Electrons that come from the $b \to u$ transition are subtracted from
the unfolded electron energy spectrum, defined later. We model the electron energy
spectrum from the  $B \to X_u l \nu$ using the De
Fazio and Neubert prescription~\cite{dfn} . The $b$-quark motion
parameters are derived in Ref. \cite{bsg}.  We scale according to the
 $B \to X_u l \nu$ branching fraction in Ref.~\cite{PDG}.

\begin{figure}[htb]
\begin{tabular}{cc}
\includegraphics[width=0.48\textwidth]{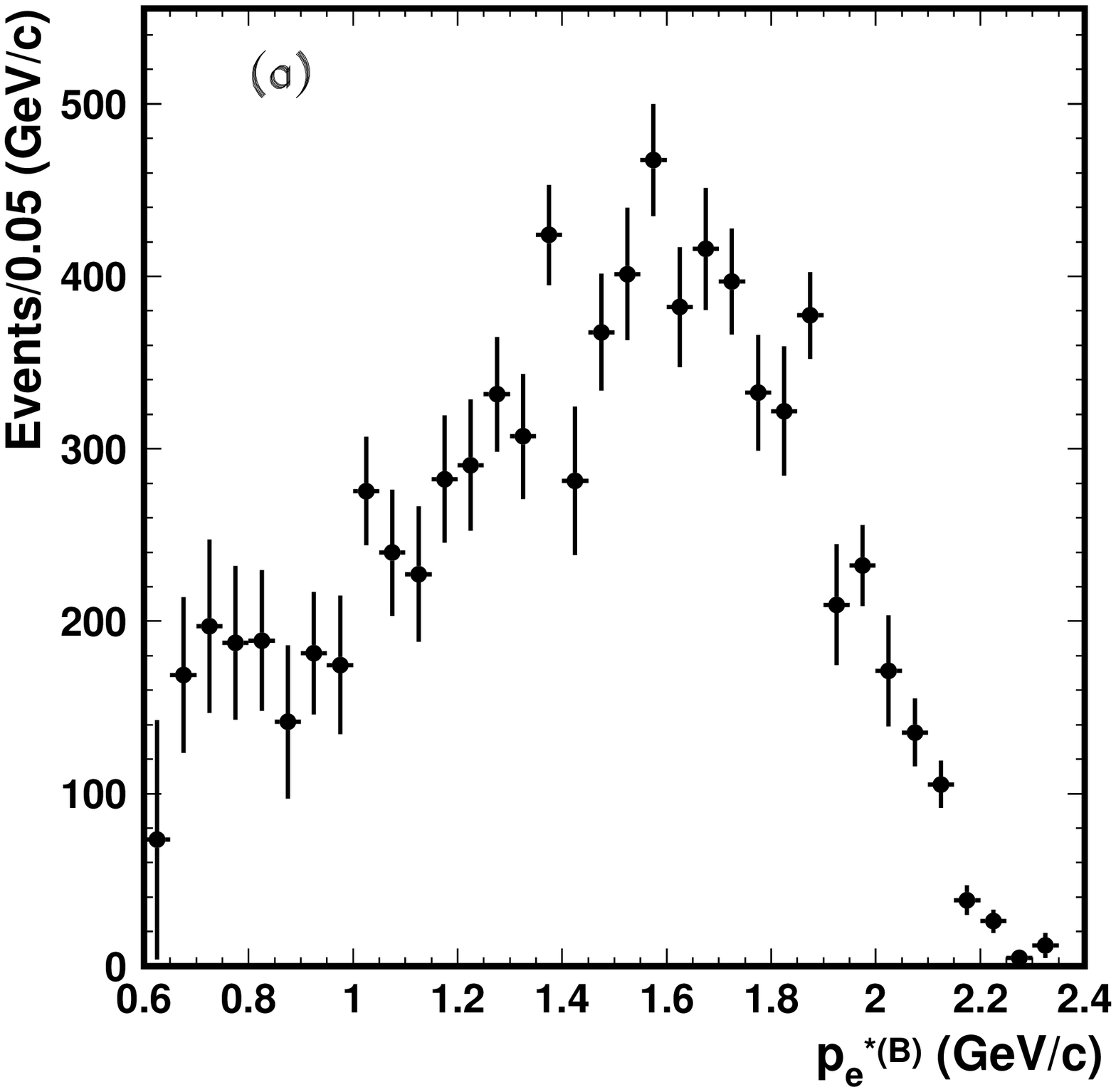}&\includegraphics[width=0.48\textwidth]{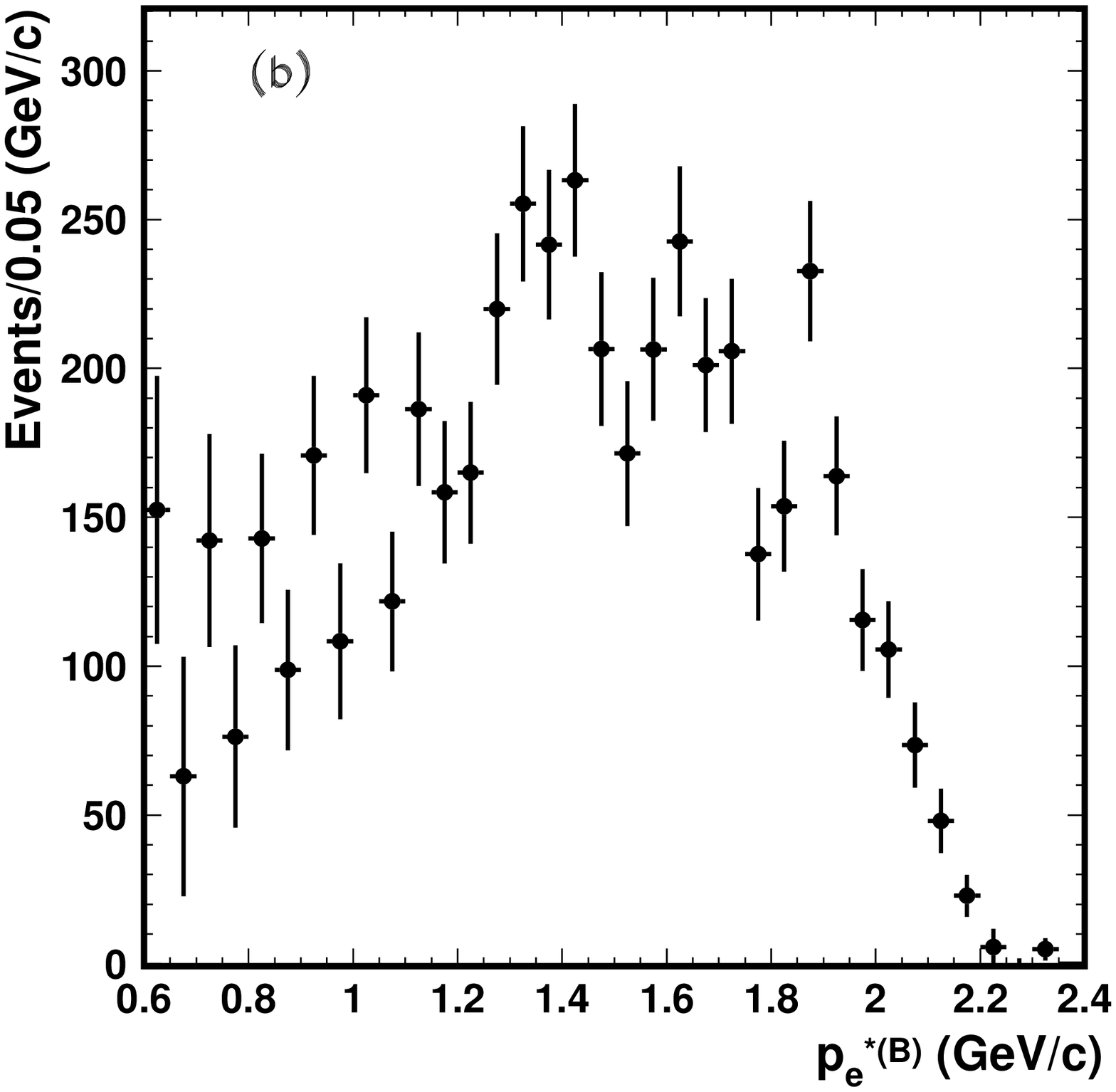}\\
\end{tabular}
\caption{The electron momentum spectrum in the $B$ rest frame after
  background subtraction, mixing and efficiency correction for
  $B^+$~(a) and $B^0$~(b).}
\label{unfold}
\end{figure}

The electron momentum spectra, after all corrections have
been applied, are shown in Figure \ref{unfold}.

\begin{table}[htb]
\caption{Number of Electrons}
\label{bzero}
\begin{tabular}
{@{\hspace{0.5cm}}l@{\hspace{0.5cm}}l@{\hspace{0.5cm}}||@{\hspace{0.5cm}}l@{\hspace{0.5cm}}||@{\hspace{0.5cm}}l@{\hspace{0.5cm}}}
\hline \hline
$B$ candidate             & $B^0$ right-sign &$B^0$ wrong-sign & $B^+$ right-sign \\  
\hline                                                                                                    
On Resonance Data           & 3367$\pm$ 58     & 1659 $\pm$ 40   & 6831 $\pm$ 83  \\   
\hline                                                                                                    
Scaled Off Resonance        & 54 $\pm$ 7       & 30  $\pm$ 6     & 106  $\pm$ 10  \\   
Combinatorial Background    & 420  $\pm$ 26    & 339 $\pm$ 29    & 1073 $\pm$ 86  \\
\hline                                                                                                    
$B \overline B$ Background  & 479 $\pm$ 21     & 856 $\pm$ 29    & 688  $\pm$ 23  \\
\hspace{5mm} Secondary      & 456 $\pm$ 20     & 847 $\pm$ 29    & 634  $\pm$ 23  \\
\hspace{5mm} Hadron Fakes   & 23  $\pm$  4     & 9   $\pm$ 2     & 54   $\pm$ 5   \\
\hline                                                                                                    
Background Subtracted       & 2414 $\pm$ 67    & 434 $\pm$ 58    & 4964 $\pm$ 122 \\
After Efficiency Correction & 4066 $\pm$ 115   & 727 $\pm$ 100   & 8371 $\pm$ 209    \\
After Mixing Correction     & 5054 $\pm$ 145   & N/A             & N/A\\
\hline \hline
\end{tabular}
\end{table}

\section{Unfolding the Electron Energy Spectrum}
To measure the first and second electron moments we need to determine the true electron energy spectrum.  
The background subtracted energy spectrum is distorted by various detector effects.
The true electron energy spectrum is extracted by performing an unfolding procedure based on 
the Singular Value Decomposition (SVD) algorithm of A.H\"ocker and V.\ Kartvelishvili~\cite{ref:13}. 
 
 \begin{figure}[htb]
\includegraphics[width=0.48\textwidth]{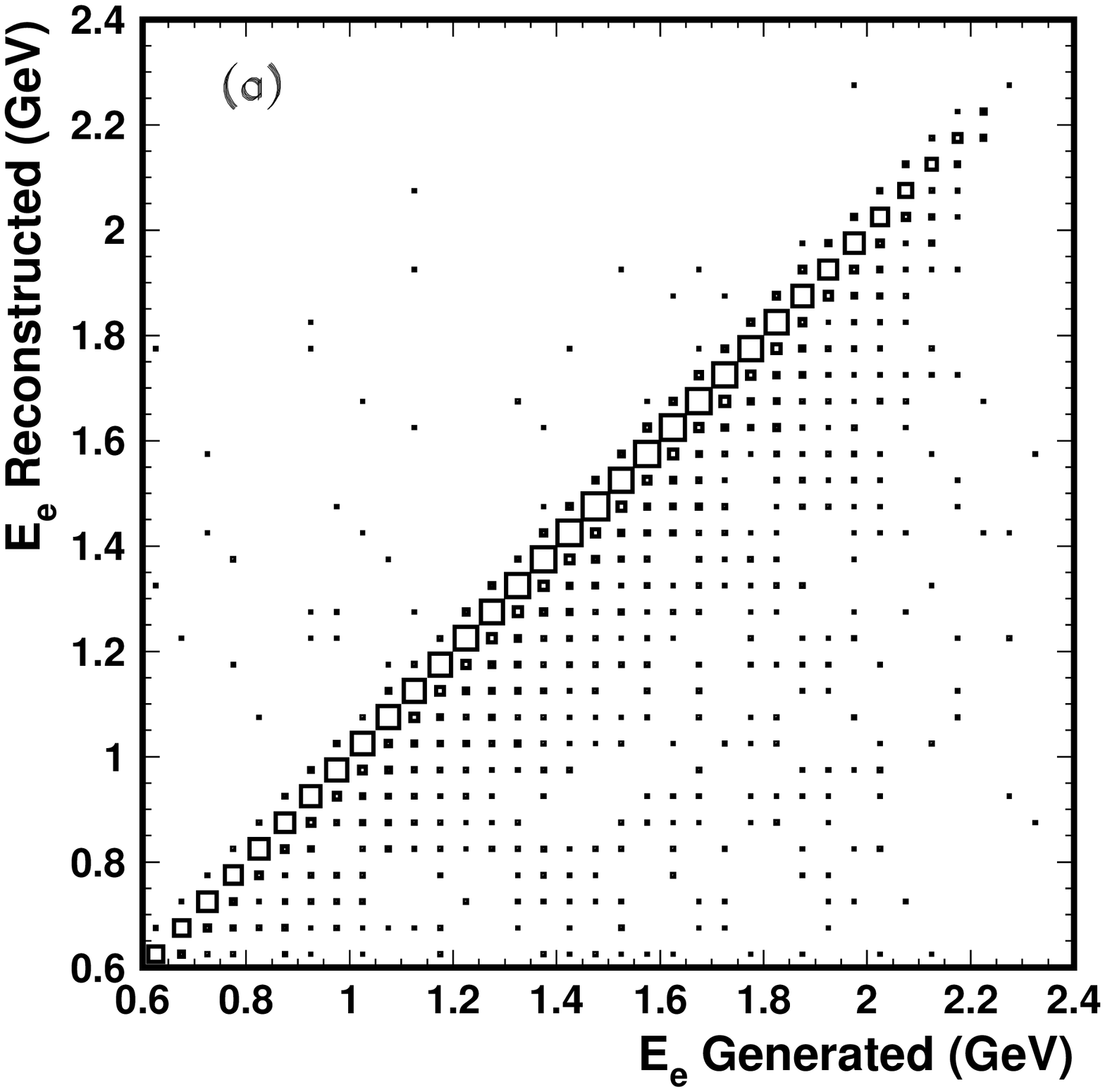}
\includegraphics[width=0.48\textwidth]{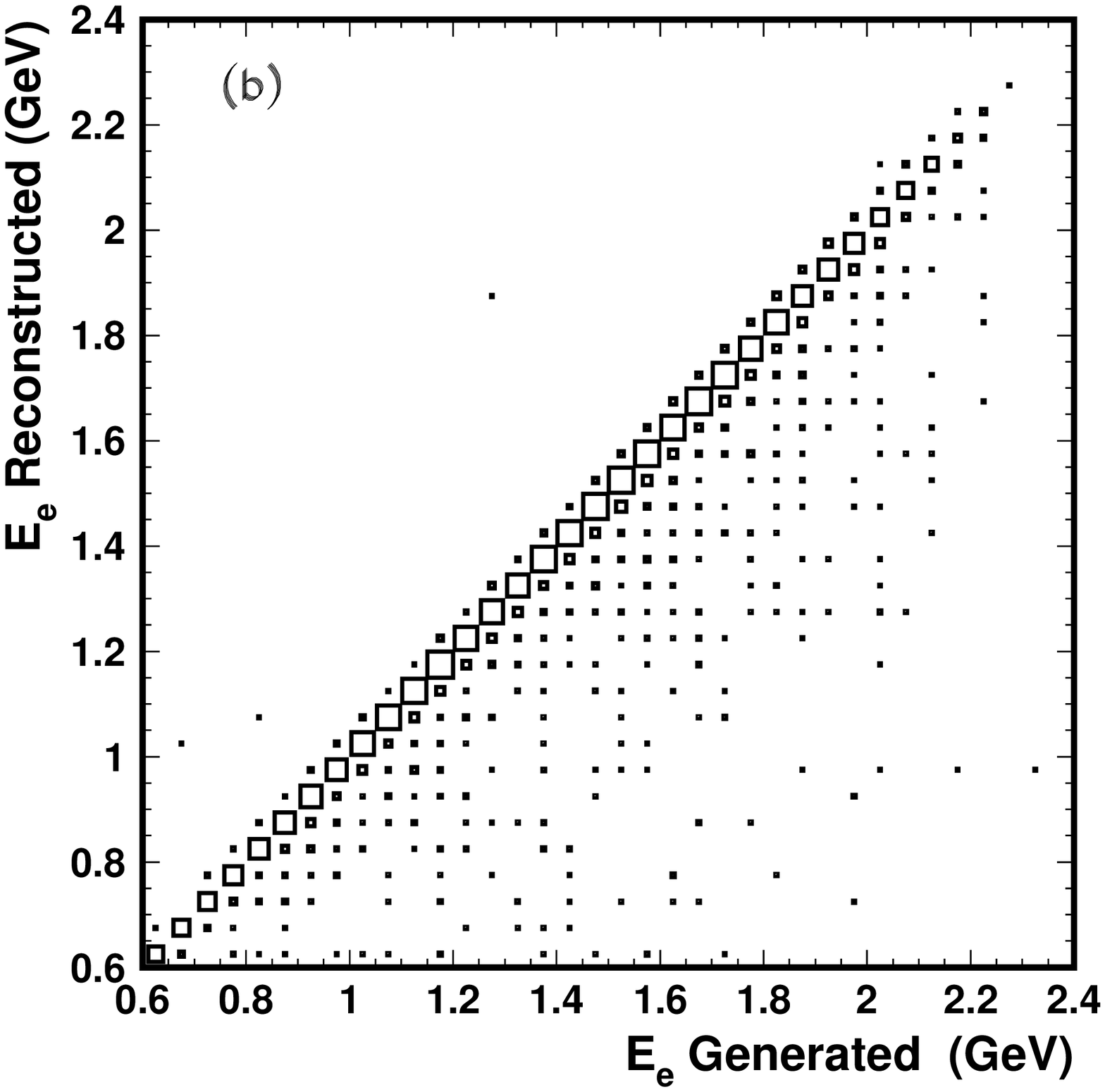}
\caption{Correlation between reconstructed and generated electron
  energy (i.e. reconstructed versus generated electron energy
  spectrum) for $B^+$~(a) and $B^0$~(b).  The high population in the
  diagonal
  bins demonstrates the high resolution in the unfolding procedure.}
\label{deconv1}
\end{figure}

Figure \ref{deconv1} shows the correlation between the reconstructed and generated electron energy. Due to the good resolution of the Belle detector, the reconstructed electron energy is highly correlated with the generated one. The off$-$diagonal bins are filled due to bremstrahlung occurring in the material in front of the calorimeter. It is important to note, for the  reliability of the unfolding, that the phase space covered by the generated energy spectrum is very similar to the reconstructed spectrum.

\begin{figure}[htb]
  \includegraphics[width=0.48\textwidth]{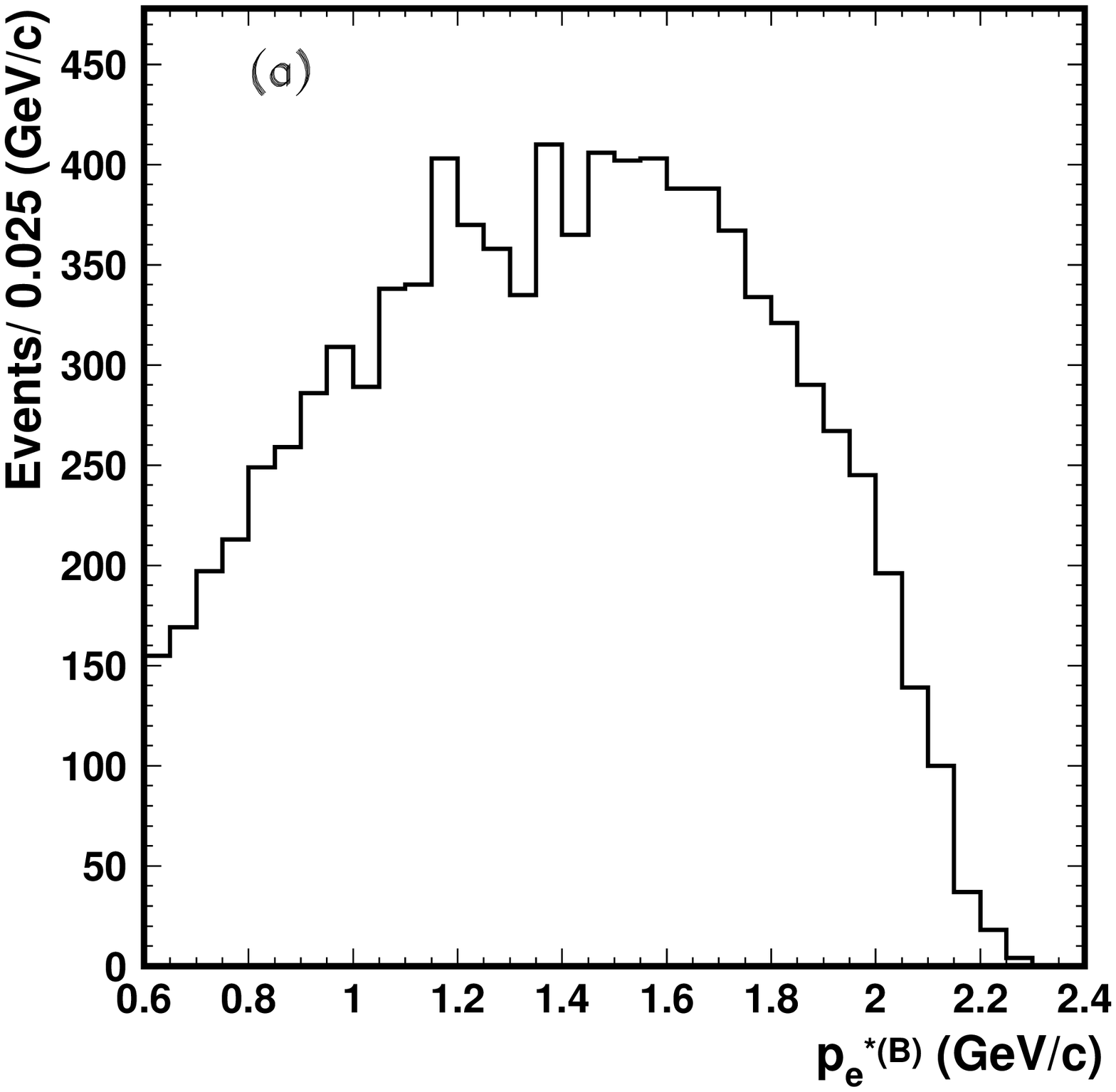}
  \includegraphics[width=0.48\textwidth]{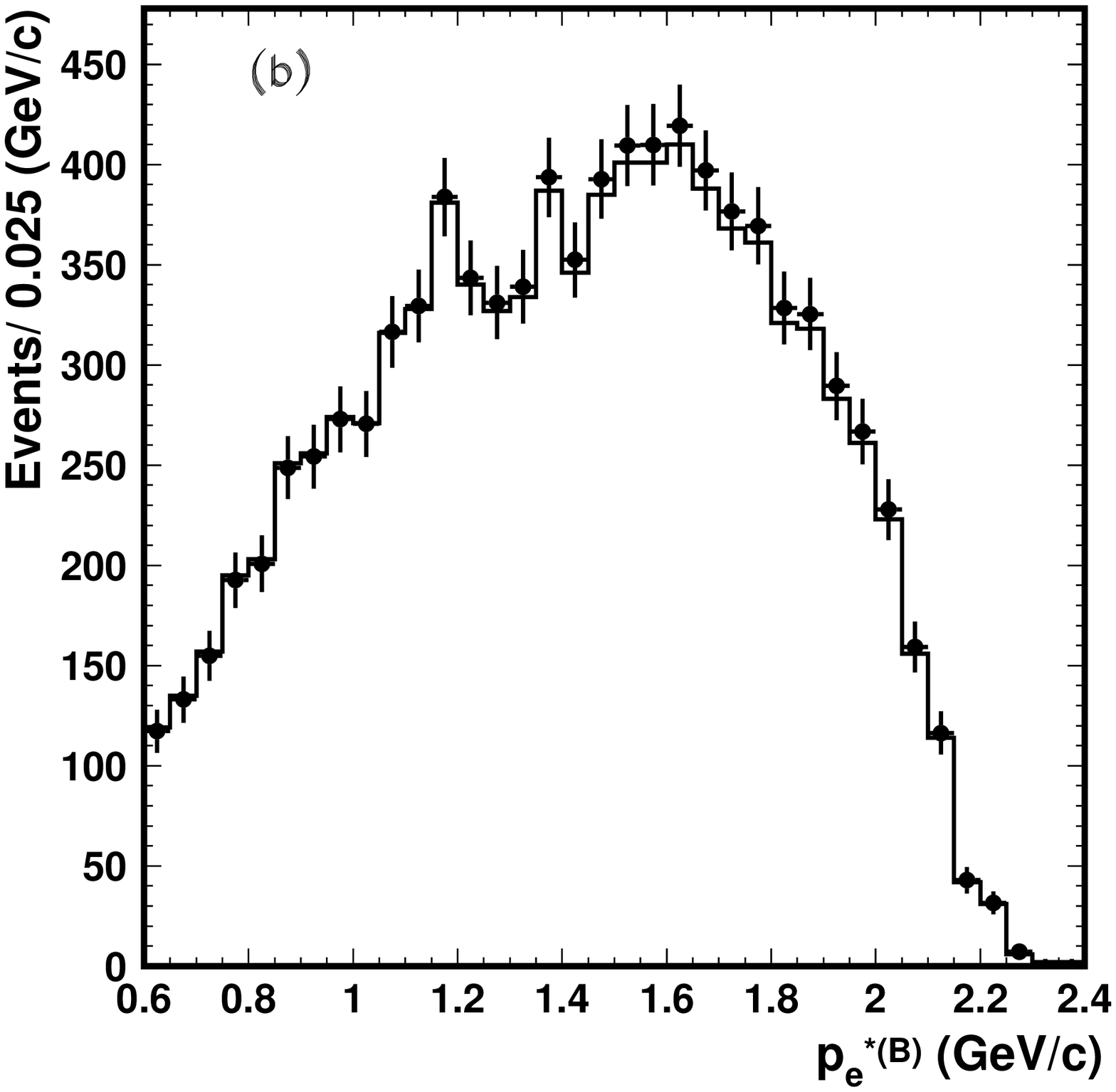}
  \caption{Generated and unfolded $E_e$ distributions for Monte Carlo
  simulated events (b).  The reconstructed
  electron spectrum after the detector simulation is show in (a).
  The continuous line in (b) is the true (generator level) $E_e$
  distribution.  The unfolded distribution is overlaid  (data
  points).}
  \label{deconv}
\end{figure}

We divide the generated energy in slices of 0.025 GeV. For each slice
 we produce a histogram of the corresponding reconstructed energy spectrum, normalised to one. The histogram has a bin width of 0.025 GeV. Bin and slicing widths have been chosen to have less than 40\% correlation (cross-talk) between bins and/or slices.   
These histograms are then used to generate a response matrix. Next, we use the SVD algorithm to perform singular value (eigenvalue) analysis
 of the response matrix.  Figure \ref{deconv} compares the unfolded
 Monte Carlo electron energy spectrum compared to the true spectrum.
 
\begin{figure}[htb]
    \includegraphics[width=0.48\textwidth]{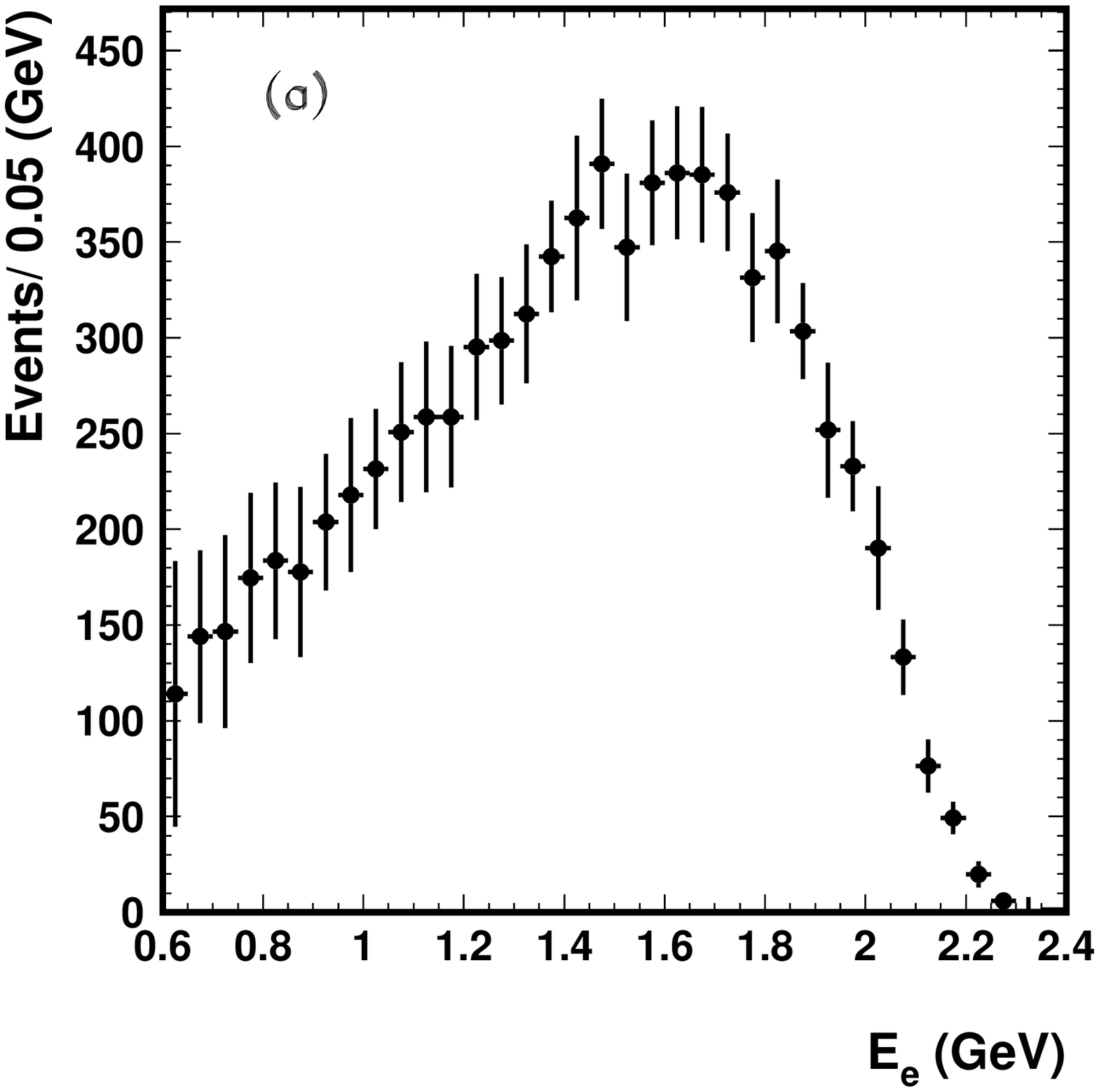}
    \includegraphics[width=0.48\textwidth]{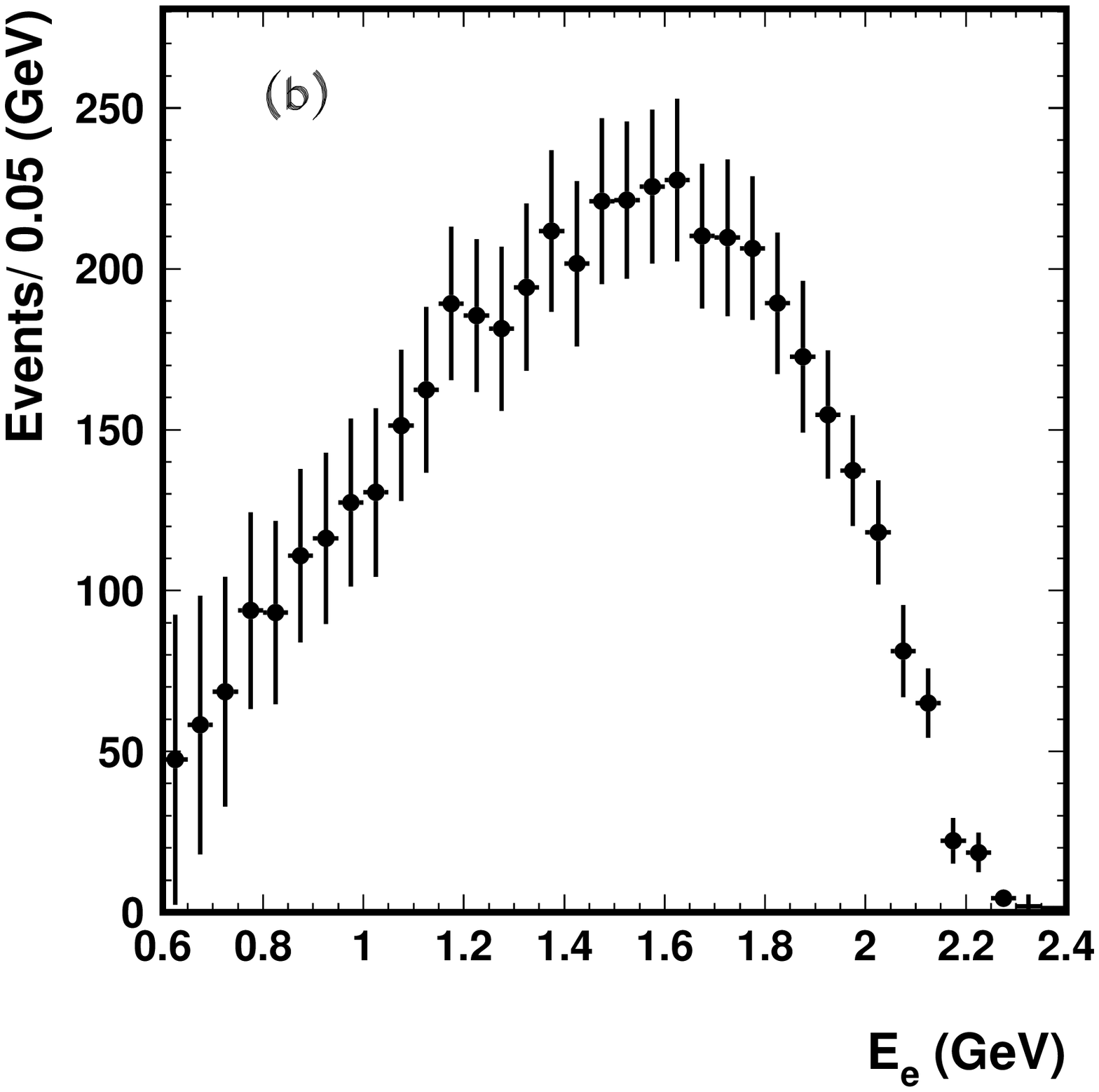}
  \caption{Unfolded $E_e$ distribution corrected for QED radiative
    effects for $B^+$~(a) and $B^0$~(b). The errors shown are statistical.}
  \label{spectrum}
\end{figure}

We weight the Monte Carlo spectra with the QED radiative effects in
 order to calculate the correct unfolding matrix.  When we unfold the
 data we remove the QED radiative effects, as the OPE does not have an
 $\mathcal{O}(\alpha)$ QED correction.  The unfolded electron energy
 spectrum for data is shown in Figure \ref{spectrum}.
 
\section{Moments of the Electron Energy Spectrum}

The first moment of the electron energy spectrum is defined to be
$M_1^e(E_{\mathrm{cut}}) = \langle E_e \rangle _{E_e>E_{\mathrm{cut}}}$ and the second
$M_2^e(E_{\mathrm{cut}})= \langle(E_e -M_1^e(E_{\mathrm{cut}}))^2\rangle_{E_e>E_{\mathrm{cut}}}$.
We measure the first and second moments with five electron energy
threshold cuts ($E_{\mathrm{cut}}=$ 0.6, 0.8, 1.0, 1.2 and 1.5 GeV) for the
$B^+$ and $B^0$ mesons.  Table \ref{moments2} gives the
final measurements of the moments, with radiative corrections applied.

\begin{figure}[htb]
  \includegraphics[width=0.48\textwidth]{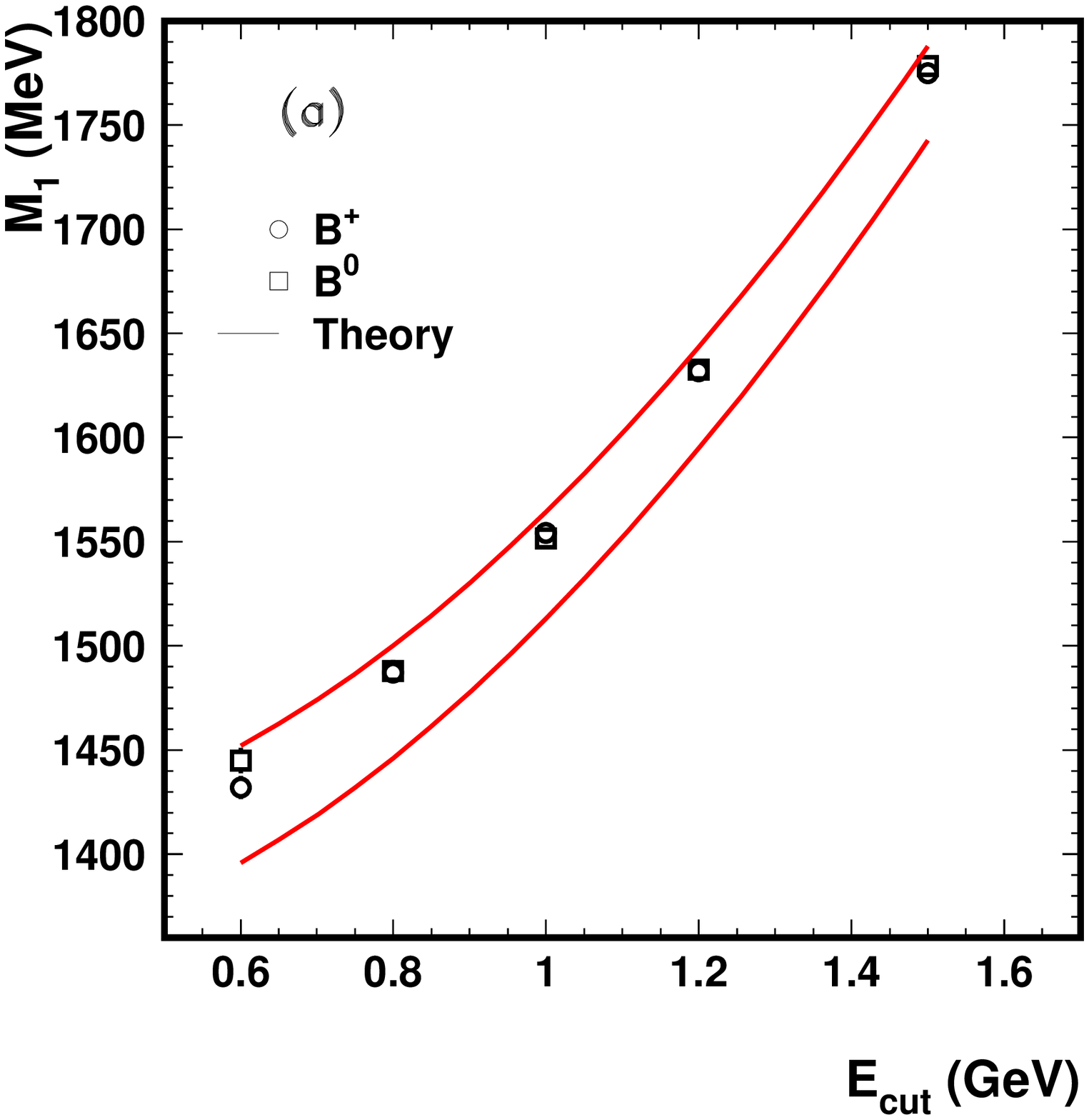}
  \includegraphics[width=0.48\textwidth]{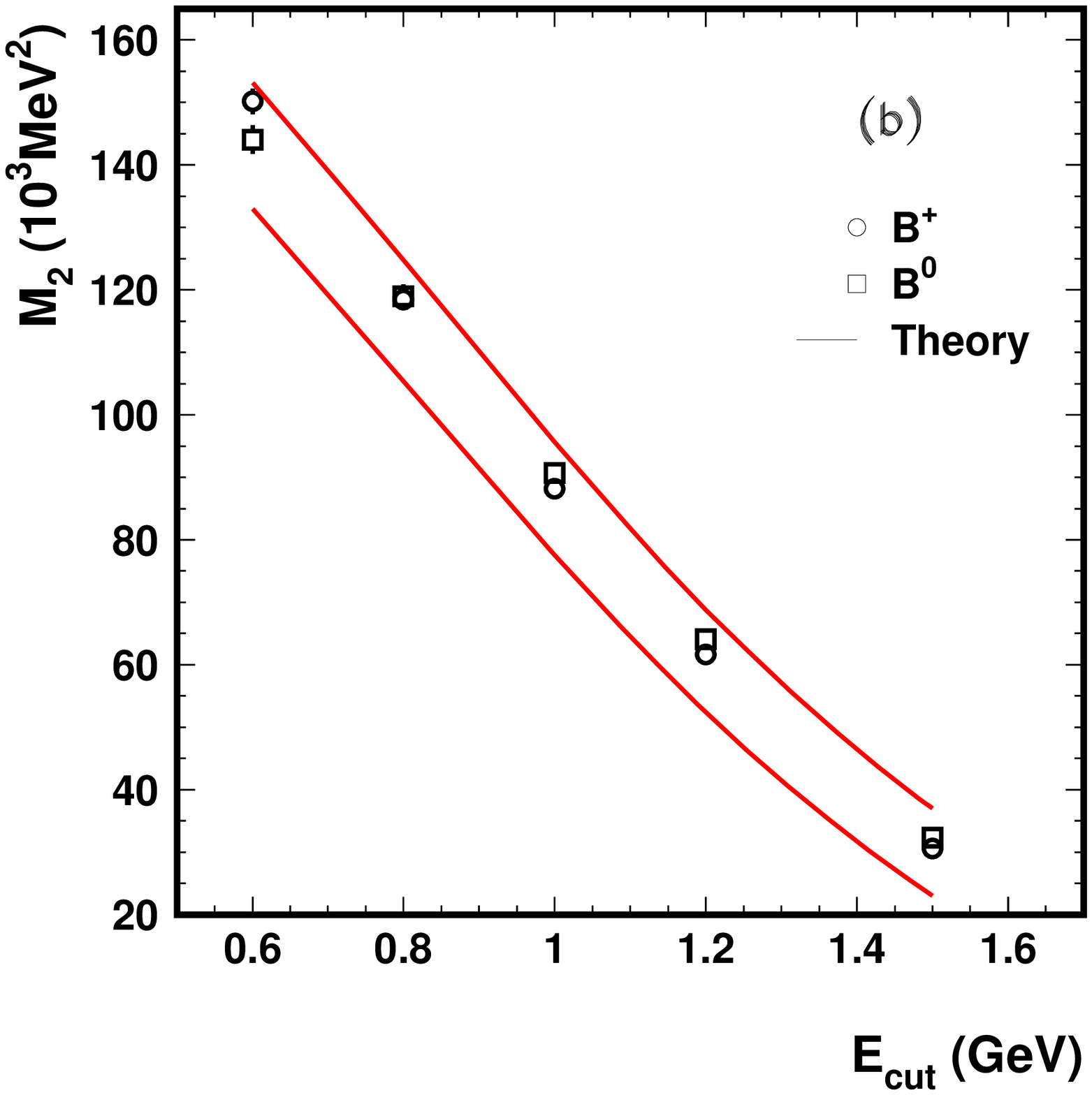}
\caption{The first and second electron energy moments, $M_1$ (a)
  and $M_2$ (b), as a function of cutoff energy $E_{\mathrm{cut}}$ with the bounds on the theory
  predictions shown as curved lines. The errors shown are statistical and
  systematic.}
\label{momentplotsep}
\end{figure}

Figure \ref{momentplotsep} shows the measured moments for the $B^0$
and $B^+$ data samples with the
theoretical bounds as a function of the threshold
electron energy.  The theoretical bounds in Figure~\ref{momentplotsep}
are taken from Ref.~\cite{ref:1}, in the kinetic mass scheme.
  

\section{Systematic Uncertainties}
The systematic uncertainty in the moments stems from event
selection, electron identification, background estimation, model
dependence, and mixing.

The uncertainty due to mis-tagging in the $B^0$ and $B^+$ samples is
derived from the magnitude of the combinatorial
background and is estimated to be~$\pm$2\%.  Uncertainty in the measured luminosity
provides an uncertainty on the continuum electron yield of $\pm$1\%.

The effect of veto cuts has been calculated as the quadratic sum
of averaged variations in the moments when selection criteria are
loosened and tightened about their chosen positions.

Subtraction of $b \to u $ decays acquires an uncertainty
associated with the renormalisation to the full inclusive
spectra.  The renormalisation factor is varied by $\pm$50\%.

The uncertainty in secondary ($B \rightarrow D \rightarrow e$) decays
derives from uncertainty in the value of the branching fractions.
Model dependence is estimated from the observed change
in the moments when the $B \to D e \nu$ and $B \to D^{*} e \nu$ decay
shape parameters
are varied.


The uncertainty in the mixing probability $\chi_d$ is based on the current
quoted error in Ref. \cite{PDG}.

In addition, we calculate systematic uncertainty associated with the magnitude of the hadron fake
contribution, electron tracking efficiency, and electron detection efficiency.

The total systematic error is obtained by adding each contribution in
quadrature.  It is important to note that the systematic errors are
limited by Monte Carlo statistics, most significantly in the $B^0$ sample.
The contributions to the systematic error are summarised in table
\ref{sysbplus} ($B^+$), and table \ref{sysbzero} ($B^0$).

\begin{table}[htb]
\caption{Breakdown of the systematic errors for the moments, $M_1$, $M_2$ for $B^+ \rightarrow X_c e \nu$ in the $B$ meson rest frame for 3 values of $E_{\mathrm{cut}}$}
\label{sysbplus}
\begin{tabular}
{@{\hspace{0.1cm}}c@{\hspace{0.1cm}}|@{\hspace{0.6cm}}c@{\hspace{0.6cm}}c@{\hspace{0.6cm}}c@{\hspace{0.3cm}}c@{\hspace{0.1cm}}c@{\hspace{0.1cm}}c@{\hspace{0.1cm}}c@{\hspace{0.1cm}}c@{\hspace{0.1cm}}
}
\hline \hline
&$M_1$&$M_1$&$M_1$&$M_2$&$M_2$ & $M_2$ \\
&[MeV]&[MeV]&[MeV]&[$10^3\rm MeV^2$]&[$10^3\rm MeV^2$]&[$10^3\rm MeV^2$]\\
$E_{\mathrm{cut}}$[GeV] & 0.6 & 1.0  & 1.5 & 0.6 & 1.0 & 1.5\\
\hline
conversions and Dalitz decays &  0.02 & 0.12 & 0.04 & 0.07  & 0.00 & 0.00 \\
e from $J/\psi$ or $\psi(2S)$ &  0.91 & 0.67 & 0.63 & 0.32  & 0.15 & 0.06 \\
\hline
$B \to D_{(s)}^{(*)} \to e$ & 1.89 & 0.43 & 0.02 & 0.65  & 0.11 & 0.00 \\
$D^{(*)}$  &  0.02 & 0.01 & 0.00 & 0.00 &  0.00 & 0.00  \\
$b \to u$ subtraction & 0.17 & 0.18 & 0.19 & 0.09 & 0.08 & 0.06   \\
\hline
continuum background &  0.00 & 0.00 & 0.00 & 0.00 & 0.00 & 0.00 \\
combinatorial background &  1.58 & 0.57 & 0.01 & 0.42 & 0.10 & 0.00 \\
\hspace{3mm} hadron fakes &  1.24 & 0.20 & 0.01 & 0.47 & 0.06 & 0.00    \\
\hline
electron detection  &  0.01 & 0.01 & 0.01 & 0.00 & 0.01 & 0.00 \\
tracking efficiency  &  0.00 & 0.0 & 0.0 & 0.0 & 0.0 & 0.0   \\
\hline \hline
total systematics & 3.59 & 1.13 & 0.67 & 1.21 &0.27  &0.09   \\
\hline\hline
\end{tabular}
\end{table}

\begin{table}[htb]
\caption{Breakdown of the systematic errors for the moments, $M_1$, $M_2$ for $B^0 \rightarrow X_c e \nu$ in the $B$ meson rest frame for 3 values of $E_{\mathrm{cut}}$}
\label{sysbzero}
\begin{tabular}
{@{\hspace{0.1cm}}c@{\hspace{0.1cm}}|@{\hspace{0.6cm}}c@{\hspace{0.6cm}}c@{\hspace{0.6cm}}c@{\hspace{0.3cm}}c@{\hspace{0.1cm}}c@{\hspace{0.1cm}}c@{\hspace{0.1cm}}c@{\hspace{0.1cm}}c@{\hspace{0.1cm}}
}
\hline \hline
&$M_1$&$M_1$&$M_1$&$M_2$&$M_2$ & $M_2$ \\
&[MeV]&[MeV]&[MeV]&[$10^3\rm MeV^2$]&[$10^3\rm MeV^2$]&[$10^3\rm MeV^2$]\\
$E_{\mathrm{cut}}$[GeV] & 0.6 & 1.0  & 1.5 & 0.6 & 1.0  & 1.5\\
\hline
conversions and Dalitz decays & 0.02  & 0.26 & 0.17 &  0.21 &  0.06 & 0.01     \\
e from $J/\psi$ or $\psi(2S)$ & 0.40  & 0.25 & 0.51 &  0.45 &  0.20 & 0.01 \\
\hline
$B \to D_{(s)}^{(*)} \to e$ &  2.38   &  0.77 & 0.05 & 0.71 & 0.13 & 0.00  \\
$\chi$ &  0.02 & 0.00 & 0.01 & 0.01 & 0.01 & 0.00    \\
$D^{(*)}$  &  0.18 & 0.03 & 0.00 & 0.07 & 0.01 & 0.00  \\
$b \to u$ subtraction &0.38&0.37&0.37&0.18&0.16& 0.12   \\
\hline
continuum background &  0.00 & 0.00 & 0.00 & 0.00 & 0.00 & 0.00 \\
combinatorial background &  0.85 & 0.31 & 0.14 & 0.20 & 0.00 & 0.02 \\
\hspace{3mm} hadron fakes &  0.93 & 0.17 & 0.01 & 0.34 & 0.03& 0.00    \\
\hline
electron detection  &  0.12 & 0.01 & 0.01 & 0.06 & 0.01 & 0.00 \\
tracking efficiency  &  0.06 & 0.00 & 0.01 & 0.03 & 0.00 & 0.01   \\
\hline \hline
total systematics & 2.76 & 0.99 & 0.67 & 0.97 & 0.30 & 0.12    \\
\hline\hline
\end{tabular}
\end{table}


\begin{table}[htb]
\caption{Measured moments, $M_1$, $M_2$ for $B \rightarrow X_c e \nu$ in the $B$ meson rest frame for five cutoff energies $E_{\mathrm{cut}}$.  The first error is the statistical, and the second error is the systematic.  The moments are corrected for QED radiative effects (using the PHOTOS package).}
\label{moments2}
\begin{tabular}
{@{\hspace{0.1cm}}c@{\hspace{0.2cm}}|@{\hspace{0.4cm}}c@{\hspace{0.4cm}}c@{\hspace{0.4cm}}c@{\hspace{0.4cm}}c@{\hspace{0.4cm}}c@{\hspace{0.4cm}}c@{\hspace{0.4cm}}c@{\hspace{0.4cm}}c@{\hspace{0.4cm}}
}
\hline \hline
$E_{\mathrm{cut}}$[GeV] & $M_1$[MeV] & $M_1$[$\rm MeV$] &
$M_2$[$10^3\rm MeV^2$] & $M_2$[$10^3\rm MeV^2$] \\
& $B^+$ & $B^0$ & $B^+$ & $B^0$\\
\hline\hline
0.6 & 1432.1 $\pm$ 4.3 $\pm$ 3.6 & 1444.9 $\pm$ 5.5 $\pm$ 2.8 & 150.1 $\pm$ 1.8 $\pm$ 1.2 & 144.0 $\pm$ 2.1 $\pm$ 1.0\\
0.8 & 1487.2 $\pm$ 3.9 $\pm$ 2.2 & 1488.0 $\pm$ 5.1 $\pm$ 1.8 & 118.4 $\pm$ 1.4 $\pm$ 0.7 & 119.0 $\pm$ 1.8 $\pm$ 0.6\\
1.0 & 1554.1 $\pm$ 3.6 $\pm$ 1.1 & 1551.5 $\pm$ 4.7 $\pm$ 1.0 & 88.1  $\pm$ 1.1 $\pm$ 0.3 & 90.7 $\pm$ 1.4 $\pm$ 0.3\\
1.2 & 1631.7 $\pm$ 3.3 $\pm$ 0.7 & 1632.6 $\pm$ 4.3 $\pm$ 0.8 & 61.7  $\pm$ 0.8 $\pm$ 0.1 & 64.1 $\pm$ 1.1 $\pm$ 0.2\\
1.5 & 1774.8 $\pm$ 2.8 $\pm$ 0.7 & 1778.2 $\pm$ 3.8 $\pm$ 0.7 & 30.6  $\pm$ 0.5 $\pm$ 0.1 & 32.3 $\pm$ 0.7 $\pm$ 0.1\\
\hline\hline
\end{tabular}
\end{table}

\section{Summary}
We report a measurement of the electron energy spectrum of the
inclusive decay $B \rightarrow X_c e \nu$ and its first and second
moments for threshold energies from 0.6 GeV to 1.5 GeV.  This set of moments,
combined with the measurements of the semileptonic branching fraction
and the moments of the hadronic  mass distribution, will be used for
the determination of the HQE parameters and of $|V_{cb}|$.

\section*{Acknowledgments}
We thank the KEKB group for the excellent operation of the
accelerator, the KEK Cryogenics group for the efficient
operation of the solenoid, and the KEK computer group and
the National Institute of Informatics for valuable computing
and Super-SINET network support. We acknowledge support from
the Ministry of Education, Culture, Sports, Science, and
Technology of Japan and the Japan Society for the Promotion
of Science; the Australian Research Council and the
Australian Department of Education, Science and Training;
the National Science Foundation of China under contract
No.~10175071; the Department of Science and Technology of
India; the BK21 program of the Ministry of Education of
Korea and the CHEP SRC program of the Korea Science and
Engineering Foundation; the Polish State Committee for
Scientific Research under contract No.~2P03B 01324; the
Ministry of Science and Technology of the Russian
Federation; the Ministry of Education, Science and Sport of
the Republic of Slovenia; the National Science Council and
the Ministry of Education of Taiwan; and the U.S.\
Department of Energy.


%


\begin{thebibliography}{99}
\bibitem{CC}
Throughout this paper, 
the inclusion of the charge conjugate mode decay is implied
unless otherwise stated.
\end{thebibliography}
\end{document}